\pdfoutput=1

\documentclass[a4paper]{spie}  

\usepackage[english]{babel}
\usepackage[utf8]{inputenc}
\usepackage[T1]{fontenc}

\usepackage{amsmath}

\usepackage{amsfonts}
\usepackage{amssymb}
\usepackage{mathrsfs}	
\usepackage{siunitx}
\usepackage{gensymb}
				
\usepackage{array}
\usepackage{color}
\usepackage{xcolor}

\usepackage{url} 
\usepackage{caption}								
\usepackage{graphicx}
\usepackage{caption}
\usepackage{fancyhdr}
\usepackage{subfigure}
\usepackage{float}
\usepackage{cite}

\title{Mid-IR AGPMs for ELT applications} 

\author{Brunella Carlomagno\supit{a}, Christian Delacroix\supit{a}, Olivier Absil\supit{a}, Pontus Forsberg\supit{b}, Serge Habraken\supit{a}, A\"issa Jolivet\supit{a}, Mikael Karlsson\supit{b}, Dimitri Mawet\supit{c}, Pierre Piron\supit{a}, Jean Surdej\supit{a} and Ernesto Vargas Catalan\supit{b} 
\skiplinehalf
\supit{a}Department of Astrophysics, Geophysics and Oceanography, University of Li\`ege, 17 all\'ee du Six
Ao\^ut, B-4000 Sart Tilman, Belgium; \\
\supit{b}{\AA}ngstr\"{o}m Laboratory, Uppsala University, L\"{a}gerhyddsv\"{a}gen 1, SE-751 21 Uppsala, Sweden; \\
\supit{c}European Southern Observatory, Alonso de C\'ordova 3107, Vitacura 7630355, Santiago, Chile
}

\authorinfo{Further author information: brunella.carlomagno@ulg.ac.be; phone: +32 43669739; ago.ulg.ac.be}

\begin{document} 
  \maketitle 

\begin{abstract}
The mid-infrared region is well suited for exoplanet detection thanks to the reduced contrast between the planet and its host star with respect to the visible and near-infrared wavelength regimes. This contrast may be further improved with Vector Vortex Coronagraphs (VVCs), which allow us to cancel the starlight. One flavour of the VVC is the AGPM (Annular Groove Phase Mask), which adds the interesting properties of subwavelength gratings (achromaticity, robustness) to the already known properties of the VVC. In this paper, we present the optimized designs, as well as the expected performances of mid-IR AGPMs etched onto synthetic diamond substrates, which are considered for the E-ELT/METIS instrument. 
\end{abstract}



\section{INTRODUCTION}
\label{sec:intro}  

Direct detection of exoplanets presents two important challenges: the presence of the extremely brighter host star and the small angular separation between the latter and its companion. However, disentangling the signal of an exoplanet from its star allows us to obtain its spectrum, which can be analyzed to infer the atmospheric composition and eventually search for biosignatures. A simple way to alleviate the contrast problem is to observe in the mid-infrared ($3-13\mu m$) regime, where the contrast between the star and the planet is reduced. In order to improve this contrast, a coronagraphic technique can also be implemented. A mask stops the light of the star, while allowing the light of the companion (as the reflected or the emitted light of the planet) to pass through the optical system\cite{Lyot}. While an amplitude mask blocks directly the light, preventing from observing regions close to the star, the phase mask rejects the starlight by destructive interferences, using (spatially-distributed) phase shifts. In particular, the vortex phase mask produces a continuous helical phase ramp, with a singularity at the center, which creates an optical vortex and hence nulls the light locally. One flavour of the vortex phase mask is the $\textit{vector vortex}$\cite{Mawet2009,Mawet11}, which consists in a rotationally symmetric halfwave plate (HWP), with an optical axis orientation that rotates about the center.

\begin{figure}[t!]
\centering
\includegraphics[width=15cm]{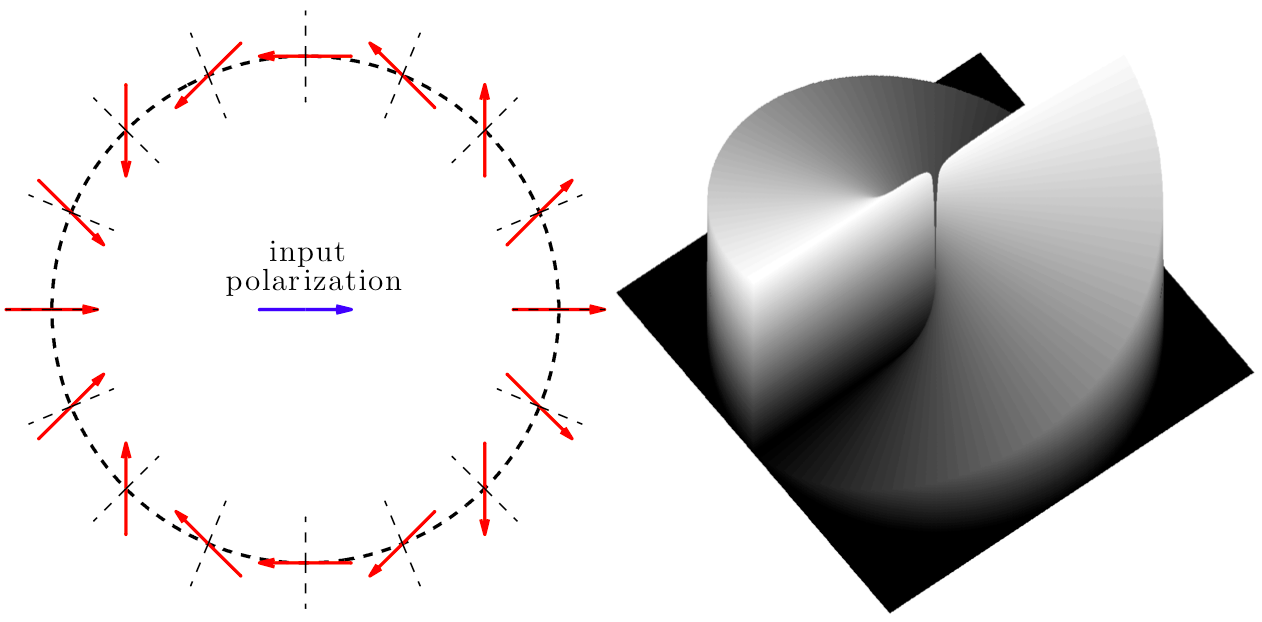}
\caption{\textbf{Left}: Illustration of the spatial variation of the optical axis orientation of a vector vortex coronagraph (VVC) of topological charge 2, obtained with a rotationally symmetric HWP. \textbf{Right}: Computed $2\times2\pi$ phase ramp created by a charge-2 vortex.}
\label{pola_phase}
\end{figure}

The effect of this vector vortex is to make the incoming horizontal polarization turn twice as fast as the azimuthal coordinate. When the vector vortex has completed the rotation with a phase ramp of $2\times2\pi$, the produced optical vortex is of charge 2 (see Fig. \ref{pola_phase}). These systems are called Vector Vortex Coronagraphs (VVCs). They present interesting properties: a small inner working angle (down to $0.9\lambda/D$), high throughput, clear off-axis 360$\degree$ discovery space and simplicity. One type of VVC is the Annular Groove Phase Mask (AGPM), which can be rendered achromatic over an appreciable spectral range thanks to the use of subwavelength gratings\cite{Mawet05,Delacroix12a}. This technology would be well suited for applications at mid-infrared wavelengths on future extremely large telescopes (ELTs).  

\section{DESIGN OPTIMIzation PROCEDURE} \label{sec:design}

\subsection{Grating parameters} \label{gratparam}

\begin{figure}[h!]
\centering
\includegraphics[width=16cm]{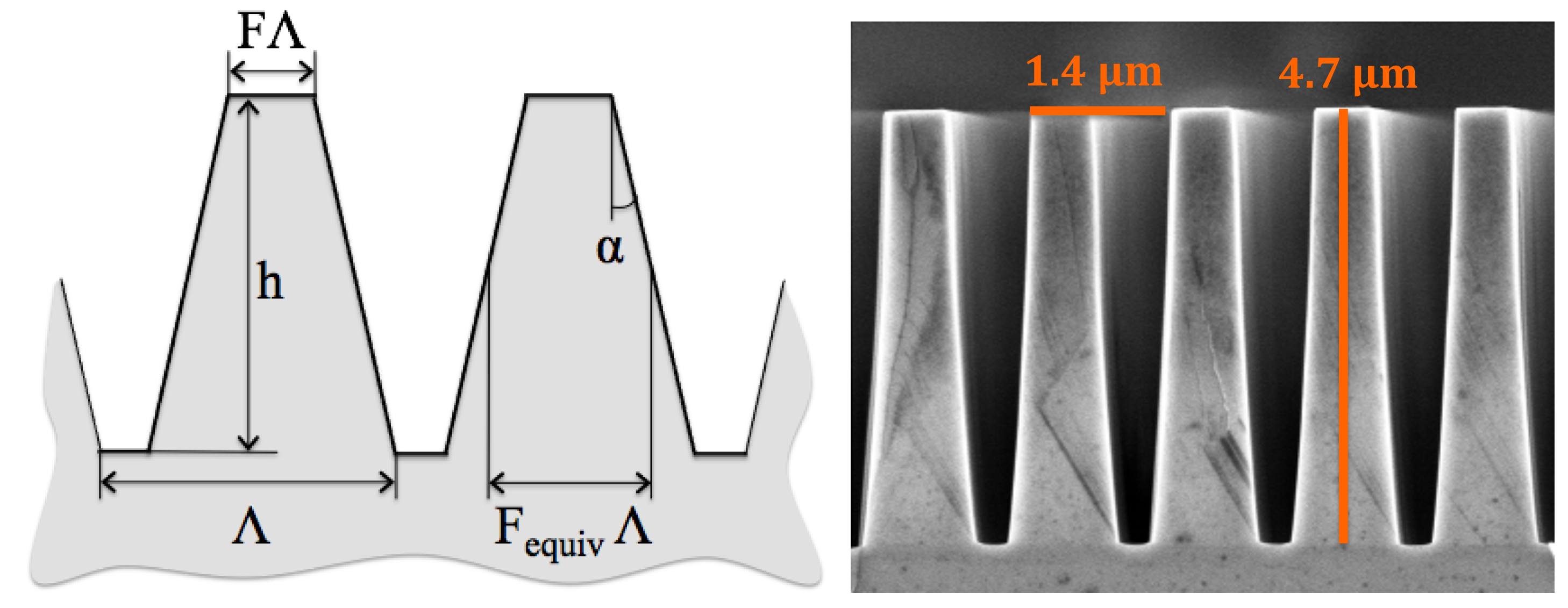}
\caption{\textbf{Left}: Schematic diagram of a trapezoidal grating. The filling factor $F$ is such that $F\Lambda$ corresponds to the line width on top of the walls. \textbf{Right}: Cross sectional view of a diamond AGPM dedicated to the L band. The grating sidewalls have an angle $\alpha \sim 3\degree$ and an average width $F_{\rm{equiv}}\Lambda \sim 0.5 \mu m$.}
\label{grating}
\end{figure}

The design optimization technique uses an algorithm based on the $\textit{Rigorous Coupled Wave Analysis}$ (RCWA)\cite{Moharam82} and coded in MATLAB\textregistered \cite{Mawet2005}. The RCWA solves Maxwell's equations without simplifying assumptions. The principal optimization parameters of the subwavelength grating profile are the filling factor $F$ and the depth $h$ (see Fig. \ref{grating}, left). 

Thanks to its great optical, mechanical, thermal and chemical characteristics, synthetic diamond turns out to be a very well suited material to manufacture such AGPMs, based on an advanced micro-fabrication technique using nano-imprint lithography and reactive ion etching\cite{Forsberg13,Forsberg14}. Due to fabrication issues, the walls of the etched grooves are not perfectly vertical, showing a slope $\alpha \sim 3 \degree$ in the present case (see Fig. \ref{grating}, right).  

\newpage
The period $\Lambda$ of the subwavelength grating is kept constant during the optimization. Its value is determined by the subwavelength limit:
\begin{equation}
\Lambda < \frac{\lambda_{\rm{min}}}{n(\lambda_{\rm{min}})}   
\end{equation}  
where $\lambda_{\rm{min}}$ is the shortest wavelength of the considered band (with a $100nm$ security margin), and $n(\lambda_{\rm{min}})$ is the refractive index of the substrate (diamond in our case) calculated at this wavelength. The refractive index was computed considering polynomial regressions and Sellmeier equations. The final values of the refractive index and the period are shown in Table \ref{refractive_period}, for several considered bands. \\

\begin{table} [h!]
\centering
\caption{Calculated diamond refractive indices and subwavelength grating periods, for several considered mid-IR spectral windows.}
\begin{tabular}{|c|c|c|c|c|}
 \hline 
  Band & Bandwidth [$\mu m$]& $\lambda_{\rm{min}} [\mu m$] &  $n(\lambda_{\rm{min}})$ & Period $\Lambda$ [$\mu m$] \\ 
 \hline 
 L & 3.5 -- 4.1 & 3.4 & 2.3814 & 1.42\\ 
 \hline 
 M & 4.6 -- 5 & 4.5 & 2.3810 & 1.89 \\ 
 \hline 
 lower N & 8 -- 11.3 & 7.9 & 2.3806 & 3.32 \\ 
\hline 
 upper N & 11 -- 13.2 & 10.9 & 2.3805 & 4.58 \\ 
  \hline 
 \end{tabular} 
\label{refractive_period}
  \end{table} 

\subsection{Null depth definition} \label{nulldef}

Theoretically, a vortex coronagraph should provide a perfect on-axis light cancellation, but imperfections prevent it. The metric used for the optimization is the $\textit{null depth}$ $N(\lambda)$, defined as the contrast, integrated over the whole point spread function (PSF)\footnote{This null depth is equal to the peak-to-peak attenuation if  the coronagraph is only limited by chromatism (in which case the coronagraph PSF is just a scaled-down version of the original PSF).}  
\begin{equation}
N(\lambda)_{\rm{theo}} = \frac{I_{\rm{coro}}(\lambda)}{I_{\rm{off}}(\lambda)} = \frac{[1-\sqrt{q(\lambda)}]^2 + \epsilon(\lambda)^2 \sqrt{q(\lambda)}}{[1+\sqrt{q(\lambda)}]^2}
\end{equation}
where $I_{\rm{coro}}$ is the signal intensity when the input beam is centered on the mask, while $I_{\rm{off}}$ is the signal intensity when the input beam is far from the mask center, $\epsilon (\lambda)$ is the phase error with respect to $\pi$, and $q(\lambda)$ is the flux ratio between the polarization components transverse electric (TE) and transverse magnetic (TM), respectively. It is in this equation that all geometrical parameters (filling factor $F$, depth $h$ and sidewall angle $\alpha$) are taken into account, via $\epsilon(\lambda)$ and $q(\lambda)$.   
  
The optimization process follows a precise procedure. First, the RCWA algorithm provides a two-dimensional map of the theoretical null depth as a function of the optimization parameters (filling factor $F$ and depth $h$), then the same algorithm is used for a more precise optimization around the optimal parameters obtained after the first step. In this paper, only a sidewall angle of 3$\degree$ has been considered. The results of the optimization are presented in Table \ref{results}.\\

\subsection{Influence of the ghost} \label{exp_perfor}

\begin{figure}[t!]
\centering
\includegraphics[width=17cm]{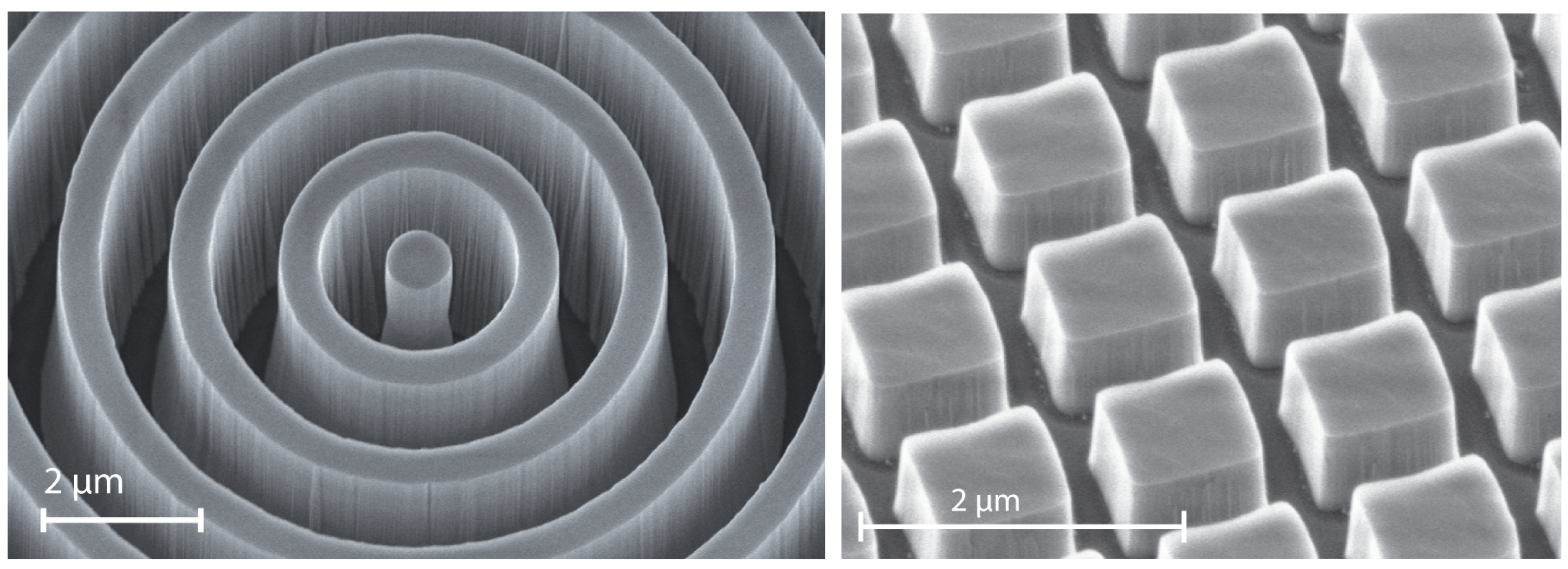}
\caption{SEM pictures of an L-band AGPM fabricated at the \AA ngstr\"om Laboratory using reactive ion etching and nanoimprint lithography. \textbf{Left}: Annular grooves etched on the frontside of the component. \textbf{Right}: Antireflective structure etched on the backside.}
\label{fabrication}
\end{figure}

The optimized parameters and null depths obtained from RCWA simulations assume that only the zeroth order of the subwavelength grating is transmitted. In practice, the presence of a ghost signal has been confirmed by laboratory measurements. It is the result of multiple incoherent reflections within the substrate. The AGPM pattern etched on the frontside of the substrate (Fig. \ref{fabrication}, left) partially reduces the reflections, acting as an antireflective layer, to some degree. Most of the ghost signal is actually caused by the reflection on the flat interface, on the other side of the substrate. Therefore, an anti-reflective grating (ARG) needs to be etched on the backside of the component (Fig. \ref{fabrication}, right) to avoid these reflections. Typically, a binary square-shaped structure is used for the antireflection. For the diamond in L band, the raw backside reflection is $17 \%$, while the ARG reduces it to $1-2\%$ \cite{Forsberg13}.

\newpage
The total null depth is then the sum of two components: 
\begin{equation}
N_{\rm{total}}(\lambda) = N_{\rm{theo}}(\lambda) + N_{\rm{ghost}}(\lambda) 
\end{equation}
where 
\begin{equation}
N_{\rm{ghost}} (\lambda)= \frac{I_{\rm{ghost}}(\lambda)}{I_{\rm{off}}(\lambda)} \; .
\end{equation}
Even though the final parameters come from a perfect optimization, where the only metric is the theoretical null depth, the expected performances for a total null depth (considering the ghost) are improved by the use of the ARG. When no ARG is present, the performances deteriorate very fast, as shown in section \ref{sec:optimal}.\\
     
\section{Simulation results and laboratory validation} \label{sec:optimaldesign}

\vskip 0.5cm
\subsection{Spectral band selection} \label{sec:select}

We have applied our optimization procedure to the mid-infrared bands L, M and N. These three bands are expected to be addressed by the METIS instrument on the future E-ELT, for which this optimization study was proposed. Our AGPM could however be used for any other high contrast imaging instrument working at these wavelengths. \\

\begin{figure}[t!]
\begin{center}
\includegraphics[width=14cm]{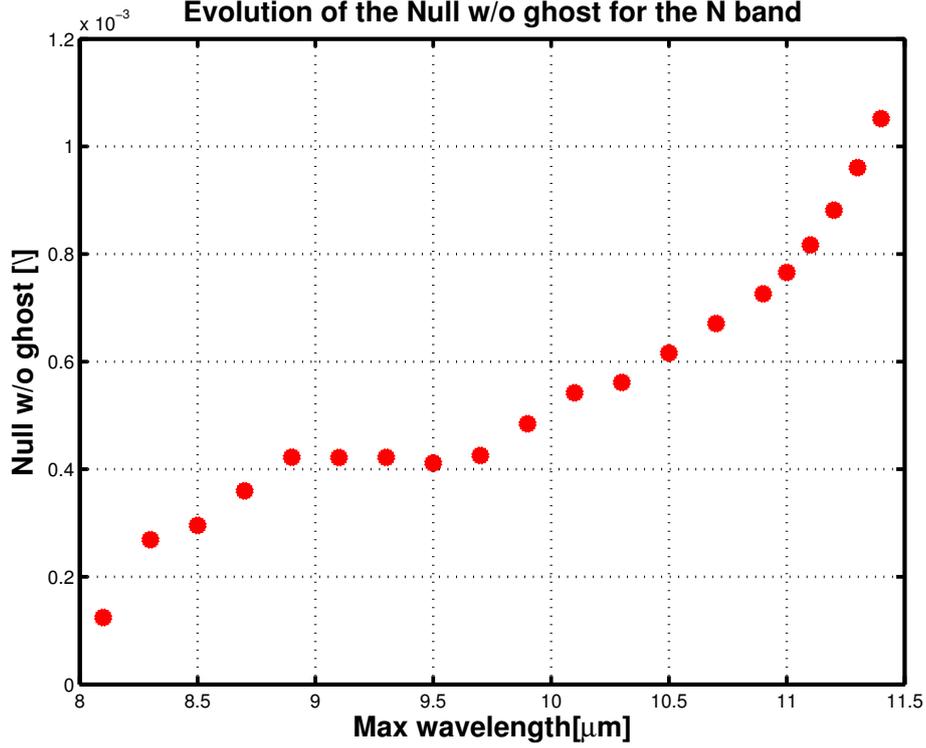}
\caption{Definition of the lower N band spectral window. Optimized mean null depth for a minimal wavelength of $8\mu m$, function of the maximal wavelength. The desired mean null depth $10^{-3}$ is obtained for the bandwidth $8-11.3 \mu m$ .}
\label{evoluzione}
\end{center}
\end{figure}

Because the N band is very wide ($8 - 13.2 \mu m$), it is difficult to cover it with a single component. Therefore, we  divided it into two sub-bands (lower and upper), on which the AGPM design was optimized. The upper N band was defined to suit the VISIR mid infrared camera of the VLT: $11-13.2 \mu m$. Recently, we have etched a few AGPMs covering this band. One of these is now installed on VISIR and shows promising performance\cite{Delacroix12b}. Concerning the lower N band, we followed a different path. A constraint was imposed on the mean null depth, which should be smaller than $10^{-3}$. We could then calculate the maximal width of the  spectral band, that we have arbitrarily started at $8 \mu m$, to define the lower N band:  $8-11.3 \mu m$ (see Fig. \ref{evoluzione}).

\newpage

\subsection{Optimal AGPM parameters and null depth at L, M and N bands} \label{sec:optimal}

The  results of our RCWA simulations are presented in Table \ref{results}, with the optimized filling factor $F$, grating depth $h$, and mean null depth $N$ over the whole bandwidth.\\

\begin{figure} [t!] 
\centering 
\includegraphics[height=19cm]{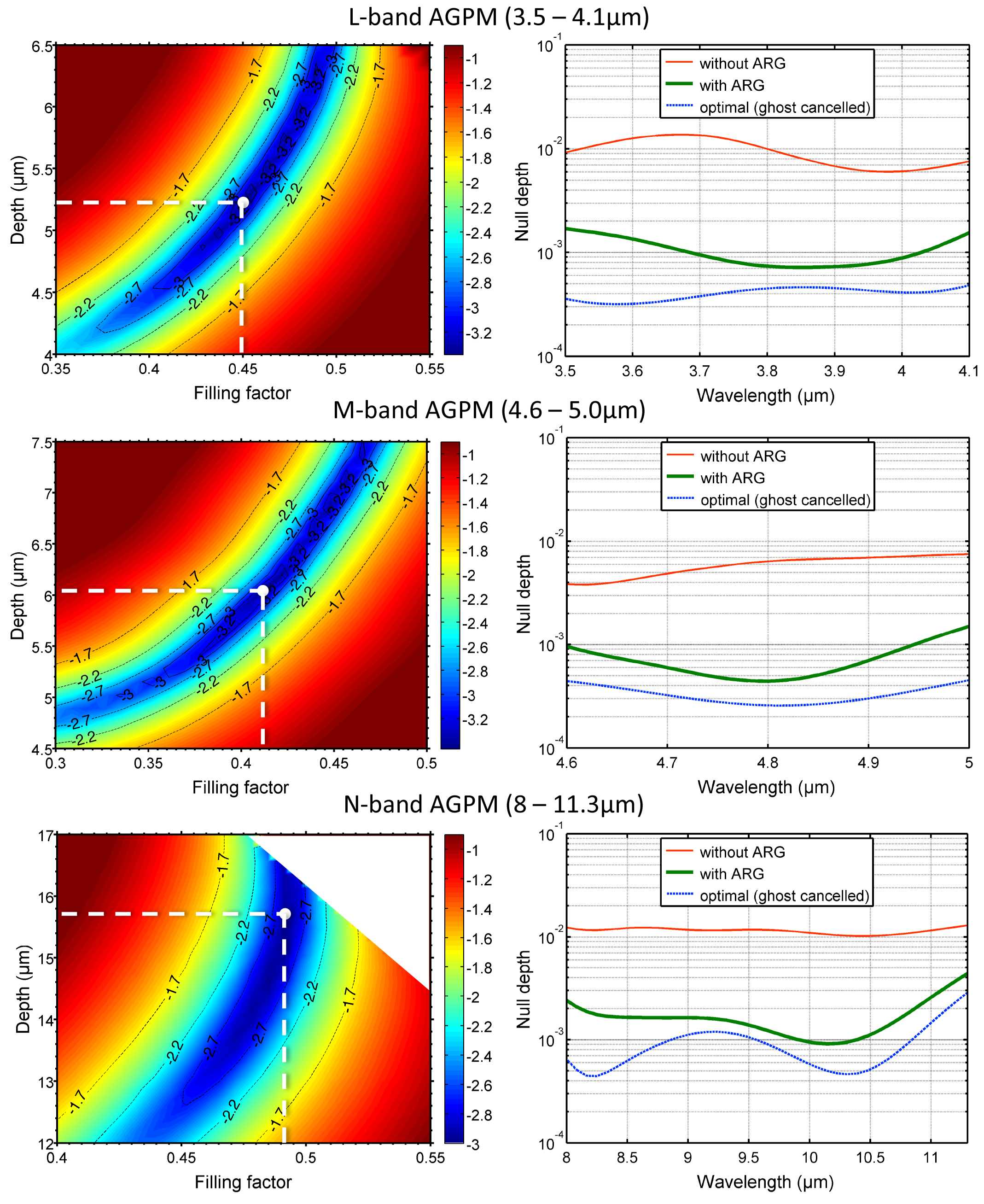}
\caption{RCWA multiparametric simulations for the L, M and N bands. \textbf{Left}: Mean null depth map, function of depth and filling factor, showing the optimal design values. For the N band, the upper right corner does not correspond to a possible geometrical solution, because of the merging of the sidewalls. \textbf{Right}: Computed coronagraphic performance of the AGPM, showing the benefits of etching an ARG on the backside of the component.}
\label{rcwa}
\end{figure}

The AGPM mean null depth (over each spectral band) is shown in Fig. \ref{rcwa} (left), function of the profile parameters $F$ and $h$.  The dark blue region in the center of the figure corresponds to the optimal parameters region, providing the best contrast expressed here on a logarithmic scale.  In Fig. \ref{rcwa} (right), we see that the theoretical optimal mean null depth $N$ is comprised between $10^{-4}$ and $10^{-3}$. When the ghost is taken into account, the somewhat degraded null depth value is still close to $10^{-3}$, showing the excellent performance of the ARG. \\

\begin{table} [h!]
\centering
\caption{Optimized filling factor $F$, grating depth $h$, and mean null depth $N$, for several considered mid-IR spectral windows.}
\begin{tabular}{|c|c|c|c|c|}
 \hline 
  Band &  Bandwidth & $F$ &   $h [\mu m]$ &  $N$\\ 
 \hline 
 L & 3.5 -- 4.1 & 0.45 & 5.22 & $4.1 \times 10^{-4}$\\ 
 \hline 
 M & 4.6 -- 5 &  0.41 & 6.07 & $3.4 \times 10^{-4}$\\ 
 \hline 
 lower N & 8 -- 11.3 & 0.49 &  15.77 & $ 10^{-3}$\\ 
 \hline 
 upper N & 11 -- 13.2 & 0.45 & 16.55 & $4.1  \times 10^{-4}$\\ 
 \hline 
 \end{tabular} 
\label{results}
  \end{table} 
  
\newpage

\subsection{L-band AGPM laboratory demonstration} \label{sec:labdemo}

To validate the performance of the L-band AGPM manufactured at the Uppsala University, we carried out laboratory coronagraphic performance tests\cite{Delacroix13} on the YACADIRE optical bench at the Observatoire de Paris. Two series of frames were recorded. The first one was obtained with the AGPM placed at the optimal position (\textit{coronagraphic frames}). The second one was obtained with the AGPM far from this position (\textit{off-axis frames}), in order to measure a reference PSF without coronagraphic effect, but still propagating into the diamond substrate. Since the presence of the AGPM slightly modifies the PSF profile near the axis, we used the \textit{raw null depth} to quantify the coronagraphic performance (rather than peak-to-peak attenuation). It is defined as the ratio between the integrated flux over a certain area around the centre of the final coronagraphic image, and the integrated flux of the same area in the off-axis image. While considering the full PSF would lead to a pessimistic result, because of all the background and high frequency artefacts, the sole central peak seems to be more appropriate. The raw null depth is then defined as:
\begin{equation}
N_{\rm{AGPM}} = \frac{\int_{0}^{\rm{FWHM}}\int_{0}^{2\pi}\tilde{I}_{\rm{coro}}(r,\theta)r\,dr\,d\theta}{\int_{0}^{\rm{FWHM}}\int_{0}^{2\pi}\tilde{I}_{\rm{off}}(r,\theta)r\,dr\,d\theta}
\end{equation}      
where we use the FWHM of the PSF as a boundary for our integral and $\tilde{I}_{\rm{coro}}$ and $\tilde{I}_{\rm{off}}$ are the medians of the reduced coronagraphic and off-axis images. The measured raw null depth was $2\times10^{-3}$ over the band, which is only within a factor 2 of the expected performance $N\sim10^{-3}$ (Fig. \ref{rcwa}, right).  Our L-band AGPMs were also recently tested and validated on sky, with two infrared cameras: VLT-NACO and LBT-LMIRCam\cite{Mawet13,Absil13,Defrere14}.\\

Finally, let us mention that no M-band AGPM has been etched so far. This will be done in the coming year and we plan to test these new AGPMs on a dedicated optical bench\cite{Jolivet14}. \\

\section{CONCLUSIONS} \label{sec:conclusions}
In this paper, we have presented the design and the expected performances of the AGPM in the mid-infrared regime, in particular the L, M and N bands. Our simulations and laboratory tests show that a null depth better than $10^{-3}$ can be achieved on most mid-infrared atmospheric windows, assuming a perfect input wavefront. This would translate into a contrast of about $5\times 10^{-6}$ at $2\lambda/D$ from the optical axis, which looks very promising for future ELT applications.

\acknowledgments     
 
The research leading to these results has received funding from the 
European Research Council under the European Union's Seventh Framework 
Programme (ERC Grant Agreement n.337569) and from the French Community 
of Belgium through an ARC grant for Concerted Research Actions.


\bibliography{article_Carlomagno_SPIE2014}   
\bibliographystyle{spiebib}   

\end{document}